\documentstyle[sprocl]{article}
\def\ra{\rightarrow}

\input{psfig}

\begin{document}
\begin{flushright}
UH-511-992-01
\\
October 2001
\end{flushright}
\vspace{.25in}
\title{CPT AND LORENTZ VIOLATIONS IN NEUTRINO OSCILLATIONS
\footnote{Presented at the Second Meeting on CPT and Lorentz Symmetry CPT'01,
August 15-18, 2001, Bloomington, Indiana.}}

\author{SANDIP PAKVASA}

\address{Department of Physics and Astronomy, \\ University of Hawaii,
\\ Honolulu, HI  96822, USA\\Email: pakvasa@phys.hawaii.edu}

%


\maketitle
\abstracts{Tests of Lorentz invariance violation and CPT Violation in neutrino
oscillations are discussed.  The sensitivity of current and future
experiments is presented.}

\section{Introduction}

We have heard about how we expect both CPT and Lorentz symmetries to be 
violated at very small scales.\cite{see}  Neutrino 
oscillations are surprisingly sensitive to CPT and Lorentz
violations and very strong constraints can be placed on symmetry
breaking parameters.  I will review the current status.

\section{CPT Violation in Neutrino Oscillations$^2$}

Consequences of $CP$, $T$ and $CPT$ violation for neutrino oscillations
have been written down before~\cite{K}. We summarize them briefly for the
$\nu_\alpha\to\nu_\beta$ flavor oscillation probabilities
$P_{\alpha\beta}$ at a distance $L$ from the source. If
\begin{equation}
P_{\alpha\beta}(L) \neq P_{\bar\alpha\bar\beta}(L) \,,
\qquad \beta \ne \alpha \,,
\end{equation}
then $CP$ is not conserved.  If
\begin{equation}
P_{\alpha\beta}(L) \neq P_{\beta\alpha}(L) \,,
\qquad \beta \ne \alpha \,,
\end{equation}
then $T$-invariance is violated.
If 
\begin{eqnarray}
P_{\alpha\beta}(L) &\neq& P_{\bar\beta\bar\alpha}(L)\,,
\qquad \beta \ne \alpha \,,
\\
\noalign{\hbox{or}}
P_{\alpha\alpha}(L) &\neq& P_{\bar\alpha\bar\alpha}(L) \,,
\end{eqnarray}
then $CPT$ is violated.
When neutrinos propagate in matter, matter effects give rise to apparent  
$CP$ and $CPT$ violation even if the mass matrix is $CP$ conserving.

The $CPT$ violating terms can be Lorentz-invariance violating (LV) or
Lorentz invariant. The Lorentz-invariance violating,
$CPT$ violating case has been discussed by Colladay and Kostelecky\cite{B}
and by Coleman and Glashow~\cite{A}. 

The effective LV $CPT$ violating interaction for neutrinos is of the form
\begin{equation}
\bar\nu_L^\alpha b_\mu^{\alpha\beta} \gamma_\mu \nu_L^\beta \,,
\label{eq:LV}
\end{equation}
where $\alpha$ and $\beta$ are flavor indices. We assume rotational
invariance in the ``preferred'' frame, in which the cosmic microwave
background radiation is isotropic (following Coleman and
Glashow~\cite{A}). The neutrino energies are
eigenvalues of

\begin{equation}
m^2/2p + b_0 \,,
\end{equation}
where $b_0$ is a hermitian matrix, hereafter labeled $b$. Consequences
of the rotational invariance violating terms {\underline b} are being investigated~\cite{kos}.

In the two-flavor case the neutrino phases may be chosen such that
$b$ is real, in which case the interaction in Eq.~(\ref{eq:LV}) is
$CPT$ odd. The survival probabilities for flavors $\alpha$ and
$\bar\alpha$ produced at $t=0$ are given by~\cite{A}
\begin{eqnarray}
P_{\alpha\alpha}(L) &=&
1 - \sin^22\Theta \sin^2(\Delta L/4)\,, \label{eq:foo}\\
\noalign{\hbox{and}}
P_{\bar\alpha\bar\alpha}(L) &=&
1 - \sin^2 2\bar\Theta \sin^2(\bar\Delta L/4) \,,\\ 
\noalign{\hbox{where}}
\Delta\sin2\Theta &=&
\left| (\delta m^2/E) \sin2\theta_m
+ 2\delta b e^{i\eta} \sin2\theta_b \right| \,,
\label{eq:delsin}\\
\Delta\cos2\Theta &=&
(\delta m^2/E) \cos2\theta_m + 2\delta b \cos2\theta_b \,.
\label{eq:delcos}
\end{eqnarray}
$\bar\Delta$ and $\bar\Theta$ are defined by similar equations with
$\delta b\to -\delta b$.  Here $\theta_m$ and $\theta_b$ define the
rotation angles that diagonalize $m^2$ and $b$, respectively, $\delta
m^2 = m_2^2 - m_1^2$ and $\delta b = b_2 - b_1$, where $m_i^2$ and $b_i$
are the respective eigenvalues. We use the convention that
$\cos2\theta_m$ and $\cos2\theta_b$ are positive and that $\delta m^2$
and $\delta b$ can have either sign.  The phase $\eta$ in
Eq.~(\ref{eq:delsin}) is the difference of the phases in the unitary
matrices that diagonalize $\delta m^2$ and $\delta b$; only one of these
two phases can be absorbed by a redefinition of the neutrino states.

Observable $CPT$-violation in the two-flavor case is a consequence of
the interference of the $\delta m^2$ terms (which are $CPT$-even) and
the LV terms in Eq.~(\ref{eq:LV}) (which are $CPT$-odd); if $\delta m^2
= 0$ or $\delta b = 0$, then there is no observable $CPT$-violating
effect in neutrino oscillations.
If $\delta m^2/E \gg 2\delta b$ then
$\Theta \simeq \theta_m$ and $\Delta \simeq \delta m^2/E$, whereas if
$\delta m^2/E \ll 2\delta b$ then $\Theta \simeq \theta_b$ and $\Delta
\simeq 2\delta b$. Hence the effective mixing angle and oscillation
wavelength can vary dramatically with $E$ for appropriate values of
$\delta b$.

We note that a $CPT$-odd resonance for neutrinos ($\sin^22\Theta = 1$)
occurs whenever $\cos2\Theta = 0$ or
\begin{equation}
(\delta m^2/E) \cos2\theta_m + 2\delta b \cos2\theta_b = 0\,;
\end{equation}
similar to the resonance due to matter effects~\cite{F,G}. The condition
for antineutrinos is the same except $\delta b$ is replaced by $-\delta
b$. The resonance occurs for neutrinos if $\delta m^2$ and $\delta b$
have the opposite sign, and for antineutrinos if they have the same
sign. A resonance can occur even when $\theta_m$ and $\theta_b$ are both
small, and for all values of $\eta$; if $\theta_m = \theta_b$, a
resonance can occur only if $\eta \ne 0$. 
If one of $\nu_\alpha$ or $\nu_\beta$ is $\nu_e$, then matter effects
have to be included. 

If $\eta=0$, then
\begin{eqnarray}
\Theta &=& \theta \,,
\label{eq:tan}\\
\Delta &=& (\delta m^2/E) + 2\delta b \,.
\label{eq:delta}
\end{eqnarray}
In this case a resonance is not
possible. The oscillation probabilities become
\begin{eqnarray}
P_{\alpha\alpha}(L) &=& 1 - \sin^2 2\theta \sin^2 \left\{ \left( {\delta m^2  
\over 4E} + {\delta b\over 2} \right) L \right\} \,,
\label{eq:P}\\
P_{\bar\alpha\bar\alpha}(L) &=& 1 - \sin^2 2\theta \sin^2 \left\{ \left(  
{\delta m^2 \over 4E} - {\delta b\over 2} \right) L \right\} \,.
\label{eq:Pbar}
\end{eqnarray}
For fixed $E$, the $\delta b$ terms act as a phase shift in the
oscillation argument; for fixed $L$, the $\delta b$ terms act as a
modification of the oscillation wavelength.

An approximate direct limit on $\delta b$ when $\alpha = \mu$ can be
obtained by noting that in atmospheric neutrino data the flux of
downward going $\nu_\mu$ is not  depleted whereas that of upward going
$\nu_\mu$~is depleted~\cite{atmos}.  Hence, the oscillation arguments in
Eqs.~(\ref{eq:P}) and (\ref{eq:Pbar}) cannot have fully developed for
downward neutrinos. Taking $|\delta b L/2| < \pi/2$ with $L\sim20$~km
for downward events leads to the upper bound $|\delta b| <
3\times10^{-20}$~GeV; the K2K results can improve this by an order of
magnitude; upward going events could in principle test
$|\delta b|$ as low as $5\times10^{-23}$~GeV.  Since the $CPT$-odd
oscillation argument depends on $L$ and the ordinary oscillation
argument on $L/E$, improved direct limits could be obtained by a
dedicated study of the energy and zenith angle dependence of the
atmospheric neutrino data.

The difference between $P_{\alpha\alpha}$ and $P_{\bar\alpha\bar\alpha}$
\begin{equation}
P_{\alpha\alpha}(L) - P_{\bar\alpha\bar\alpha}(L) =
- 2 \sin^22\theta \sin\left({\delta m^2 L\over2E}\right) \sin(\delta b L) 
\,, \label{eq:deltaP}
\end{equation}
can be used to test for $CPT$-violation. In a neutrino factory, the
ratio of $\bar\nu_\mu \to \bar\nu_\mu$ to $\nu_\mu \to \nu_\mu$ events
will differ from the standard model (or any local quantum field theory
model) value if $CPT$ is violated. Fig. 1 shows the
event ratios $N(\bar\nu_\mu \to \bar\nu_\mu)/N(\nu_\mu \to \nu_\mu)$
versus $\delta b$ for a neutrino factory with 10$^{19}$ stored muons and
a 10~kt detector at several values of stored muon energy, assuming
$\delta m^2 = 3.5\times10^{-3}$~eV$^2$ and $\sin^22\theta = 1$, as
indicated by the atmospheric neutrino data~\cite{atmos}. The lack of
equality of $\nu$ and $\bar{\nu}$ events at $\delta b=0$ comes about due
to the difference between $\nu$ and $\bar{\nu}$ cross-sections, assuming
equal fluxes.  The error bars
in Fig. 1  are representative statistical
uncertainties. The node near $\delta b = 8\times10^{-22}$~GeV is a
consequence of the fact that $P_{\alpha\alpha} =
P_{\bar\alpha\bar\alpha}$, independent of $E$, whenever $\delta b L = n
\pi$, where $n$ is any integer; the node in Fig.~1 is for
$n=1$. A $3\sigma$ $CPT$ violation effect is possible in such an
experiment for $\delta b$ as low as $3\times10^{-23}$~GeV for stored
muon energies of 20~GeV. For $\nu_e's$, if the relevant $\delta m^2$
is smaller than $10^{-10} eV^2$ (``just-so''), then for large mixing
$\delta b_{\nu_{e}}{_{\nu_x}}$ can be bounded by $10^{-27}$ GeV from
solar neutrino data.  If all
$\delta m^2$'s were sufficiently small $(<10^{-17} eV^2)$, then data from
SN1987A and future data from AGN neutrinos can probe $\delta b$ down to
$10^{-40} GeV$.

We have also checked~\cite{V} the observability of $CPT$ violation at other
distances, assuming the same neutrino factory parameters used above.
For $L=250$~km, the $\delta b L$ oscillation argument in
Eq.~(\ref{eq:deltaP}) has not fully developed and the ratio of $\bar\nu$
to $\nu$ events is still relatively close to the standard model value.
For $L=2900$~km, a $\delta b$ as low as $10^{-23}$~GeV may be observable
at the $3\sigma$ level. However, longer distances
also have matter effects that simulate $CPT$ violation, which have to be
corrected for. 

Lorentz invariant CPT violation can arise if e.g. $\delta m_{ij}^2$ and
$\theta_{ij}$ are different for neutrinos and antineutrinos.  Constraints
on such differences are rather weak~\cite{V}.  Taking advantage of this, a very
intriguing proposal has been made by Barenboim et al.~\cite{baren} 
They propose that
in the $\nu$ sector, the $\delta m^2$ and mixing are ``conventional'' and
nearly bimaximal; namely $\delta m^2_{23} \sim$ atmospheric and $\delta
m^2_{21} \sim$ LMA.  Whereas in the $\bar{\nu}$ sector $\delta m^2_{23}
\sim 0 (eV^2), \delta m^2_{21} \sim$ atmospheric and the mixing is large
in the 1-2 sector and small (of order LSND) in 2-3 sector. Then the
$\bar{\nu_\mu}-\bar{\nu}_e$ conversion in LSND~\cite{aguilar} is
accounted for, 
and the solar neutrinos are unaffected as no $\bar{\nu}'s$ are emitted in the
sun.  This proposal can be tested by Mini-Boone seeing LSND effect in
$\bar{\nu}_\mu$ beam but not in the $\nu_\mu$ beam and by the fact that the
$\nu_e$ and $\bar{\nu}_e$ oscillations with $\delta m^2_{atm}$ will be
very different (present in former and absent in latter).  For example,
KAMLAND~\cite{suzuki} will see no effect in Reactor $\bar{\nu}_e's$ even
if LMA is the correct solution for solar $\nu_e's$.

\section{Lorentz Invariance Violation in Neutrino Oscillations}

A general formalism to describe small departures from exact Lorentz
invariance has been developed by Colladay and Kostelecky~\cite{colladay}.  This
modification of Standard Model is renormalizable and preserves the gauge
symmetries.  When rotational invariance in a preferred frame is imposed,
the formalism developed by Coleman and Glashow~\cite{coleman} can be
used.  
In this form, the main effect (at high energies) of the violation of Lorentz
invariance is that each particle species $i$ has its own maximum attainable
velocity (MAV), $c_i$, in this frame.  The Lorentz violating parameters are
$c^2_i - c_j^2$.

There are many interesting consequences~\cite{coleman}:  evading of GZK  cut-off,
possibility of ``forbidden'' processes at high thresholds e.g. $\gamma \ra
e^+ + e^-, p \ra e^+ + n + \nu, \ \mu \ra \pi + \nu_\mu, \ \mu \ra  e + \gamma$
etc.  Another new allowed phenomenon is the oscillation of massless
neutrinos.

Even if neutrinos were massless, the flavor eigenstates could be
mixtures of velocity (MAV) eigenstates and the flavor survival
probability (in the two flavor case) is given by
\begin{equation}P_{\alpha \alpha} = 1- sin^2 2 \theta \sin^2
\left (\frac{\delta c}{2} LE \right )
\end{equation}
where $\delta c= c_1-c_2$.

Identical phenomenology for neutrino oscillations arises in the case of
flavor violating gravity or the violation of equivalence principle, with
$\delta \gamma \Phi$ replacing $\delta c$.  Here, $\delta \gamma =
\gamma_1-\gamma_2$ is the difference in the PPN parameters which
violates the equivalence principle and $\Phi$ is the gravitational
potential.  This test of equivalence principle was first proposed~\cite{gasperini} by
Gasperini and by Halprin and Leung.  There does not seem to be a consistent
theoretical scheme from which such consequences follow.  In other words,
a theory of gravity which would agree with the classic tests of GR and
also violate equivalence principle has not yet been
found~\cite{halprin}.  

This form of massless neutrino oscillations was very interesting at one
time.  The reason was that a single choice of parameters $\delta c$ and
$\sin^2 2 \theta$ could account for both atmospheric and solar neutrinos
with $\nu_e - \nu_\mu$ mixing~\cite{glashow}.  However, now $\nu_\mu-\nu_e$ can no
longer account for atmospheric neutrinos~\cite{kamiokande} and the LE dependence is ruled
out for atmospheric neutrinos~\cite{learned}.  A description of solar
neutrinos, even including the recent SNO data, is
still possible~\cite{ray}; with the choice of parameters:  $\delta c/2 \sim
{10^{-24}}$ and large mixing.  For $\nu_\mu-\nu_x$ mixing, the results of
NUTEV~\cite{romo} can be used to constrain $\delta c/2 < 10^{-21}$ (for $sin^2 2 \theta >
10^{-3})$, and future Long Baseline experiments~\cite{minos} will extend the bounds to
$10^{-23}$ for large mixing.

In the general case, when neutrinos are not massless, the energies are
given by 
\begin{equation}
E_i= p+ m^2_i/2p + c_i p
\end{equation}
There will be two mixing angles (even for two flavors) and the survival
probability is given by
\begin{equation}
P_{\alpha \alpha}= 1- sin^2 2 \Theta \sin^2 (\Delta L/4)
\end{equation}
where
\begin{eqnarray}
\Delta \sin 2 \Theta & = & \mid(\delta m^2/E) \sin 2 \theta_m + 2 \delta c e^{i \eta}
\ E \sin 2 \theta_c \mid , \\
\Delta \cos 2 \Theta & = & (\delta m^2/E) \cos 2 \theta_m + 2 \delta c 
\ E \cos 2 \theta_c
\end{eqnarray}

One can also write the most general expression including the CPT
violating term of Eq. (6) and even extending to three flavors.  But there
is not enough information to constraint the many new parameters.

\section{Summary}
When data from Long Baseline experiments and eventually neutrino
factories become available, CPT and Lorentz violation in neutrino
oscillations can be probed to new and significant levels.  It would be
especially useful to have detectors capable of distinguishing between
$\nu$ and $\bar{\nu}$ events.

\section*{Acknowledgments}

I thank Vernon Barger,  Alan Kostelecky and Tom Weiler for
discussions.  This work is supported in part by U.S.D.O.E. under
grant DE-FG 03-94ER40833.

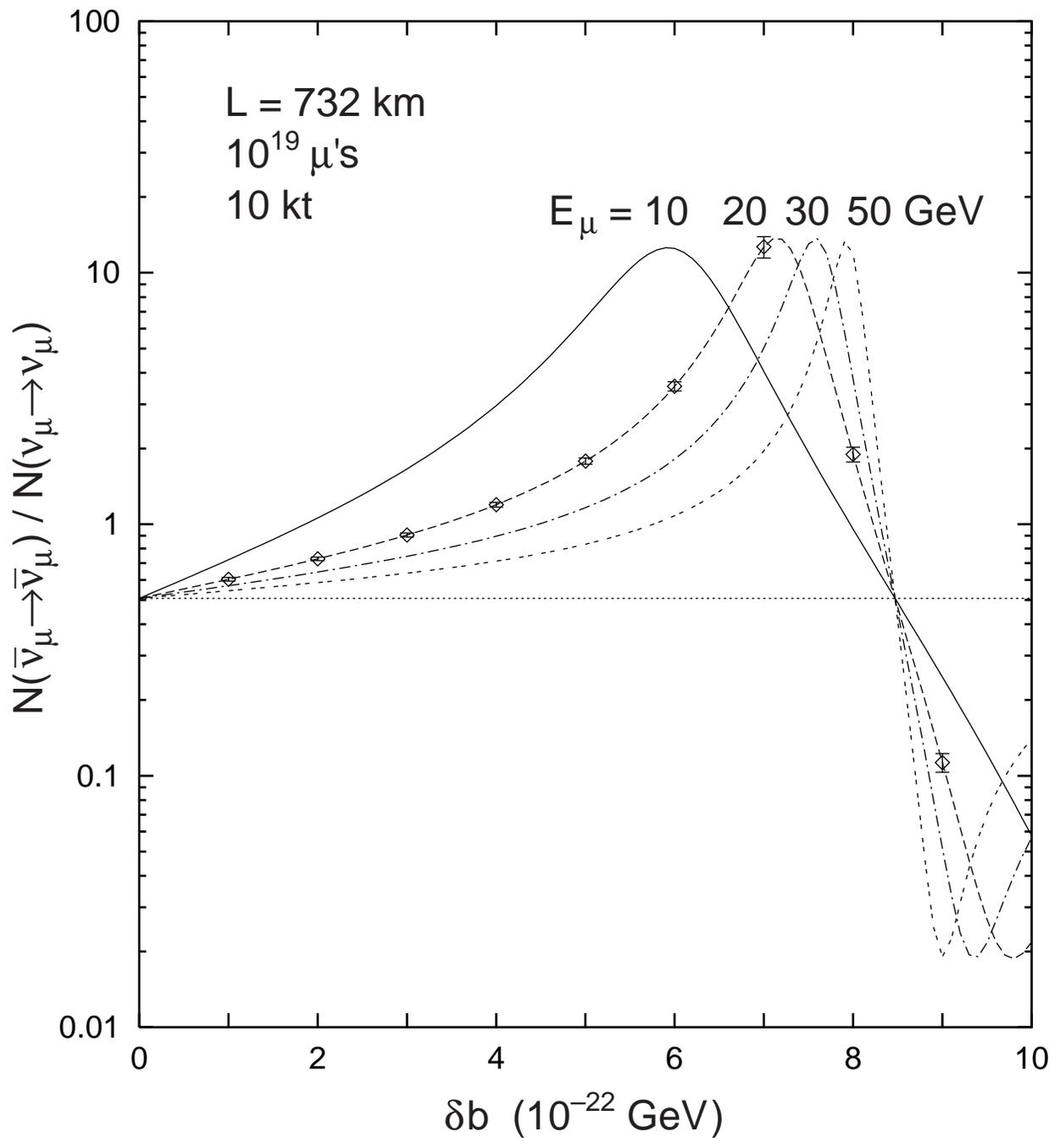
\end{document}